\documentclass[10pt]{article} 

\title{Clustered multi-state models with observation-level random effects, mover-stayer effects and dynamic covariates: Modelling transition intensities and sojourn times in a study of psoriatic arthritis }
\date{}
\author{Sean Yiu*, Vernon T. Farewell, Brian D.M. Tom\\
*Email-address: Sean.yiu@mrc-bsu.cam.ac.uk\\
MRC Biostatistics Unit, Cambridge, CB2 0SR, UK}
\usepackage{booktabs}
\usepackage{amsmath}
\usepackage{amssymb}
\usepackage{graphicx} 
\usepackage{xypic}
\usepackage{tikz}
\usepackage{fullpage}
\usepackage{rotating}
\usepackage{afterpage}
\usetikzlibrary{fit,positioning}
\usepackage{bm}
\usepackage[title,titletoc,toc]{appendix}

\newcommand{\nc}{\newcommand}
\nc{\etal}{{\it{et al. }}}
\graphicspath{{H:/Pictures/}}

\begin{document}
\maketitle
\noindent \textbf{Summary:} In psoriatic arthritis, it is important to understand the joint activity (represented by swelling and pain) and damage processes because both are related to severe physical disability. This paper aims to provide a comprehensive investigation in to both processes occurring over time, in particular their relationship, by specifying a joint multi-state model at the individual hand joint-level, which also accounts for many of their important features. As there are multiple hand joints, such an analysis will be based on the use of clustered multi-state models. Here we consider an observation-level random effects structure with dynamic covariates and allow for the possibility that a subpopulation of patients are at minimal risk of damage. Such an analysis is found to provide further understanding of the activity-damage relationship beyond that provided by previous analyses. Consideration is also given to the  modelling of mean sojourn times and jump probabilities. In particular, a novel model parameterization which allows easily interpretable covariate effects to act on these quantities is proposed.\\


\section{Introduction}
In psoriatic arthritis, manifestations of the disease typically result in joints becoming swollen and/or painful (active joints), which are reversible through treatment/management strategies or spontaneously, and may lead to permanent joint damage. The interplay between disease activity (as measured by activity in the joint) and damage is believed to be of a causal nature with a previous investigation performed by O'Keeffe \etal (2011) providing an extensive discussion on the topic. In that analysis, amongst others, individual joint-level three-state models consisting of a not active and not damaged state, active and damage state and an absorbing damaged state were proposed and produced strong evidence of a greatly increased transition rate to damage when a joint is active (compared to a joint being not active). The three-state models were fitted under a working independence assumption (the three-state processes are independent within an individual) with a robust covariance matrix used to adjust standard errors (Lee and Kim, 1998). The purpose of this paper is to extend the current modelling framework so that greater confidence with regard to the association between activity and damage can be achieved, and also to inform on other important clinical questions.\\
\indent From a statistical point of view, it is important to adjust for observed, and where possible, unobserved characteristics which are believed to be strongly related to the processes of interests, i.e. confounder variables. An analysis that does not may produce spurious associations between included covariates and the outcome. Therefore, as extensions to O'Keeffe \etal (2011), dynamic covariates (covariates which describe important aspects of previous developments of the process) are included to allow current transition intensities to depend on previous history (relaxing the Markov assumption), random effects are introduced into the transition intensities to account for unobserved heterogeneity and provide a more efficient estimation procedure, and a mover-stayer model (Frydman, 1984) is considered to allow for the possibility that some patients (stayers) have no propensity to develop damaged joints. Whilst much research has focused on the impact of disease activity on joint damage, no research has yet considered the reverse association (i.e. impact of damage on activity). Specifically, it is of interest to investigate whether the disease activity process changes with damage onset, and how if so. To inform on the possible association, the absorbing damaged state is further subdivided into an active and damaged state and a not active and damaged state, thereby allowing the disease activity process to be modelled even after a joint has become damaged. The resulting model then utilizes the entire data-set, as opposed to previously where the disease activity process was stopped once a joint had become damaged. By considering a mover-stayer model, it is also possible to investigate whether the activity process is different between movers (those who have the propensity to develop damaged joints) and stayers, and this will also contribute new knowledge towards the relationship between damage and activity.\\
\indent Clustered progressive multi-state models constructed using random effects have previously been proposed in the panel data literature. See for example, Cook \etal (2004), O'Keeffe \etal (2012) and Sutradhar and Cook (2008). In our context, these models introduce time-invariant, possibly multivariate random effects at the patient-level to account for the correlation between joints from the same patient, time-invariant unobserved heterogeneity and relaxation of the Markov assumption. A novel feature of our work is the use of observation-level multivariate random effects in clustered non-progressive multi-state models to account for correlation and time-varying unobserved heterogeneity and the introduction of dynamic covariates to explicitly relax the Markov assumption. The proposing of this random effects structure was motivated by the extensive lengths of follow-up and the possibly non-predictable changes in unobserved heterogeneity due to treatment/management strategies employed by the clinic and the spontaneous nature of joint activity. Such observations are less likely to result in unobserved heterogeneity being time-invariant or time-varying but deterministic (which is enforced by patient-level random effects). We also note that along with generalised estimating equations, copulas (Diao and Cook, 2014) and expanded state space models (Tom and Farewell, 2011) have been proposed to handle clustering. Although there are considerable advantages to such models, they are particularly difficult to formulate and implement when more than two intermittently observed non-progressive multi-state processes are of interest.\\
\indent The natural multi-state modelling parameterization allows covariates to act on transition intensities in a proportional hazards framework. Therefore easily interpretable covariate effects on these transition intensities can be obtained. Another natural way to view a multi-state process is in terms of its sojourn times (time spent in a state before a transition occurs) and jump probabilities (probability of transitioning to a state given a transition occurs). If these quantities are of interest, a model parameterization which allows easily interpretable covariate effects to act on these quantities would be useful, especially as current parametrizations may not allow for such interpretation. We consider this issue to motivate a modification of the original three-state model.\\ 
\indent The next section introduces the psoriatic arthritis data on which this analysis is based. 
\section{Psoriatic arthritis data}
Psoriatic arthritis is an inflammatory arthritis associated with the skin condition psoriasis. At the University of Toronto psoriatic arthritis clinic, over 1000 patients have been followed-up longitudinally since it began in 1978 with clinic visits scheduled $6-12$ months apart. In particular, at these visits, the active and damaged joint counts are recorded at the individual joint-level, amongst other measurements, and therefore permit statistical modelling at this level of detail. In this investigation, focus will be on the 28 hand joints (14 joints in each hand, see Figure 1 for more details), which can result in severe physical disability if active and/or damaged. Furthermore, this investigation is based on 743 patients who entered the clinic with no damage in either hand, so that patients are more comparable in their initial state of disease progression, and had greater than two clinic-visits. A dynamic covariate which requires previous observations will be constructed in the next section. Of this subset of patients, 69$\%$ (514 of 743 patients) had no damage at the end of their follow-up, which motivates consideration of  a stayer population. The mean follow-up time was 10 years and 8 months with interquartile range of 11 years and 6 months. The mean and median number of clinic visits were 12.7 and 8 respectively, and this ranged from 2 to 57. The mean and medium inter-visit times were 10 and 6 months, with standard deviation of 1 year and 3 months. At clinic entry, the mean age at arthritis onset was 36 years and 8 months with a standard deviation of 13 years and 4 months, whilst the mean arthritis duration was 5 years and 2 months with a standard deviation of 7 years and 2 months. Furthermore, $55\%$ of patients were male and $45\%$ female.\\
\indent In total, there were 264,208 observed transitions over all hand joints in the data. The observed transition matrix is as follows:
\[\bordermatrix{~ & \bar{A}\bar{D} & A\bar{D} & \bar{A}D & AD\cr
                  \bar{A}\bar{D} & 217,976 & 11,008 &623 &158 \cr
                  A\bar{D} & 12,250 & 8,599 & 200& 169 \cr
\bar{A}D & 0& 0& 10,935&680\cr
AD& 0& 0& 882& 728\cr}\]
where $\bar{A}$ and $A$ denote the absence and presence of activity in the joint respectively, and $\bar{D}$ and $D$ denote if the joint has been clinically assessed as not damaged and damaged respectively.\\
\indent The next section describes a six-state model which will be useful for jointly investigating the activity and damage processes.

\begin{figure}[h!]
\centering
\includegraphics[scale=0.5]{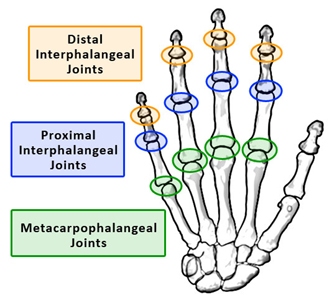}
\caption{Diagram of the type of joints in each hand. This investigation focuses on 14 joints in each hand consisting of the distal interphalangeal, proximal interphalangeal and metacarpophalangeal joints in each finger and the proximal interphalangeal and metacarpophalangeal joints in the thumb. This figure was obtained from the arthritis fact sheet on the Georgia tech web page usability.gtri.gatech.edu.}
\end{figure}

\section{Six-state model for transition intensities}
Multi-state models provide a convenient framework when the evolution of a stochastic process is of interest (Commenges, 1999 and Anderson, 2002). This investigation demonstrates their use for the joint analysis of the disease activity and damage processes occurring in each individual hand joint. Specifically, consider the following four-state representation depicted in Figure 2. 

\begin{figure}[h!]
\centering
\begin{tikzpicture}
\tikzstyle{main}=[circle,fill=white,draw=black,text=black,minimum size=1.3cm]
\tikzstyle{connect}=[->,shorten >=1pt,node distance=10cm,semithick]
  \node[main] (0)  {3 
$\bar{A}D$};
  \node[main] (1) [right=2cm of 0] {4 
$AD$};
  \node[main] (2) [below=2cmof 0] {1
$\bar{A}\bar{D}$};
\node[main] (3) [right=2cm of 2] {2
$A\bar{D}$};
\

\path (0) edge [connect, bend left] node [pos=0.5, sloped, above]{$\lambda_{34}$} (1);
  \path (1) edge [connect, bend left] node [pos=0.5, sloped, below]{$\lambda_{43}$} (0);
 \path (2) edge [connect, bend left] node [pos=0.5, left]{$\lambda_{13}$} (0);
\path (3) edge [connect, bend right] node [pos=0.5, right]{$\lambda_{24}$} (1);
  \path (3) edge [connect, bend left,above] node [pos=0.5, sloped, below]{$\lambda_{21}$} (2);
 \path (2) edge [connect, bend left] node [pos=0.5, sloped, above]{$\lambda_{12}$} (3);
\end{tikzpicture}
\caption{
Multi-state model describing the disease activity and damage processes jointly for movers.}
\end{figure}
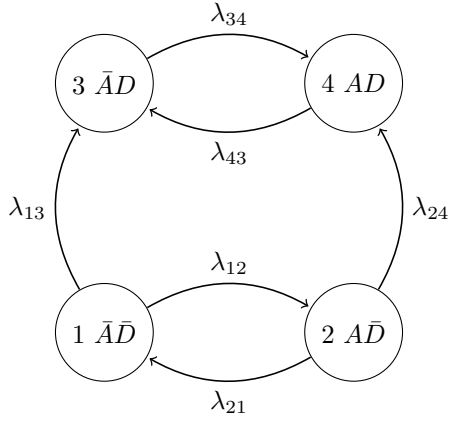

The process characteristics, particularly the reversibility of activity and permanent nature of damage respectively, are reflected in the non-zero transition intensities which describe the instantaneous rate of transitioning between states. It is implicit that this representation describes the possible transitions of movers since $\lambda_{13}$ and $\lambda_{24}>0$. If however a stayer population exists with regard to developing damaged hand joints, their disease activity process can be described by the multi-state diagram in Figure 3.

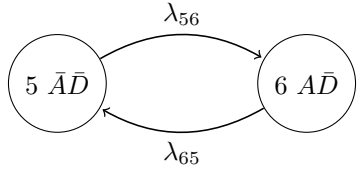
\begin{figure}[h!]
\centering
\begin{tikzpicture}
\tikzstyle{main}=[circle,fill=white,draw=black,text=black,minimum size=1.3cm]
\tikzstyle{connect}=[->,shorten >=1pt,node distance=10cm,semithick]
  \node[main] (0)  {5
$\bar{A}\bar{D}$};
  \node[main] (1) [right=2cm of 0] {6 
$A\bar{D}$};
\
\path (0) edge [connect, bend left] node [pos=0.5, sloped, above]{$\lambda_{56}$} (1);
  \path (1) edge [connect, bend left] node [pos=0.5, sloped, below]{$\lambda_{65}$} (0);
\end{tikzpicture}
\caption{
Multi-state model describing activity process for stayers.}
\end{figure}
\indent Let $\lambda^l_{rsij}$ denote the transition intensity from state $r$ to $s$ for the $l$th joint of the $i$th patient at the $j$th clinic visit (at each clinic visit all 28 hand joints are observed). In order to more simply investigate specific clinical aspects described in the introduction (and to formulate a more parsimonious model), the transition intensities are parameterized as follows:
\begin{align*}\tag{1}
\lambda^l_{34ij}&=\lambda^l_{12ij}\exp(\beta^{\bar{A}A}_{Damaged})\\
\lambda^l_{43ij}&=\lambda^l_{21ij}\exp(\beta^{A\bar{A}}_{Damaged})\\
\lambda^l_{24ij}&=\lambda^l_{13ij}\exp(\beta^{\bar{D}D}_{Active})\\
\lambda^l_{56ij}&=\lambda^l_{12ij}\exp(\beta^{\bar{A}A}_{Stayer})\\
\lambda^l_{65ij}&=\lambda^l_{21ij}\exp(\beta^{A\bar{A}}_{Stayer}).
\end{align*}
Thus regression coefficients are used to provide a simple representation of the effect of damage and mover-stayer status on activity transitions as well as activity on the damage transition. Furthermore, we let
\begin{align*}\tag{2}
\lambda^{l}_{12ij}&=\lambda^{\bar{A}A}_0\exp(\bm{\beta}^{\bar{A}A} \bm{z}^l_{ij}+u_{ij})\\
\lambda^{l}_{21ij}&=\lambda^{A\bar{A}}_0\exp(\bm{\beta}^{A\bar{A}} \bm{z}^l_{ij}+\alpha u_{ij})\\
\lambda^{l}_{13ij}&=\lambda^{\bar{D}D}_{0}\exp(\bm{\beta}^{\bar{D}D} \bm{z}^l_{ij}+v_{ij})\\
\end{align*}
where $\lambda^{\bar{A}A}_0$, $\lambda^{A\bar{A}}_0$, $\lambda^{\bar{D}D}_{0}$ are constant baseline intensities, $\bm{\beta}^{\bar{A}A}$, $\bm{\beta}^{A\bar{A}}$, $\bm{\beta}^{\bar{D}D}$ are vectors of regression coefficients, $\bm{z}^l_{ij}$ is a vector of covariates associated with the $l$th joint from the $i$th patient at the $j$th clinic visit, and $u_{ij}$, $v_{ij}$ are realizations of zero-mean bivariate normal observation-level random effects. Here $\alpha \in \mathbb{R}$ is an unknown parameter to be estimated which allows $u_{ij}$ to act differently across the different transition intensities associated with the activity process. Although not formally stated, we include time-dependent dynamic covariates in $\bm{z}^l_{ij}$ to relax the Markov assumption. Specifically, observed history of the activity process is summarized through a joint-specific covariate denoted as adjusted mean activity (AMA, Iba$\tilde{\text{n}}$ez \etal (2003), Figure 4 provides a description), whilst a patient's state of disease progression is reflected through the current number of damaged joints attained. On average, AMA was calculated as 0.093 with standard deviation of 0.19 in our data. Given these dynamic covariates, current transition intensities from the multi-state process are then assumed independent of previous process history, i.e. the Markov assumption. The random effects are assumed independent across time (with respect to $j$) and can be seen as accounting for unobserved heterogeneity not due to previous process history (where adjustments to unobserved heterogeneity related to previous history are provided through the dynamic covariates), which is still unaccounted for in the model. It is worth noting that the explicitly specified regression coefficients in (1) and (2) correspond to covariates with different modelling assumptions. The covariates in $\bm{z}^l_{ij}$ are assumed to remain constant between clinic visits and therefore are relevant when this is true (time-invariant covariates) or a reasonable approximation (for example when the covariate process is unlikely to be highly fluctuating between clinic visits). For simplicity, such covariates are also usually included if understanding the relationship between these covariates and the outcome are not of primary interest but some form of adjustment for these covariates are necessary. In contrast, the regression coefficients $\beta_{Damaged}$ and $\beta_{Active}$ are describing the effect of a binary variable representing a joint being damaged and active respectively whilst reflecting the stochastic nature of these processes, and therefore provide more realistic measures of association. This is especially useful because these are the clinical aspects of primary interest. The regression coefficient $\beta_{Stayer}$ is similar in nature to $\alpha$ as it describes the effect of a partially observable binary variable ($stayer=1$ and $mover=0$); it can only be known that patients with damage are movers and that patients with no damage are either movers or stayers.\\
\begin{figure}[h!]
\centering
\includegraphics[scale=0.5]{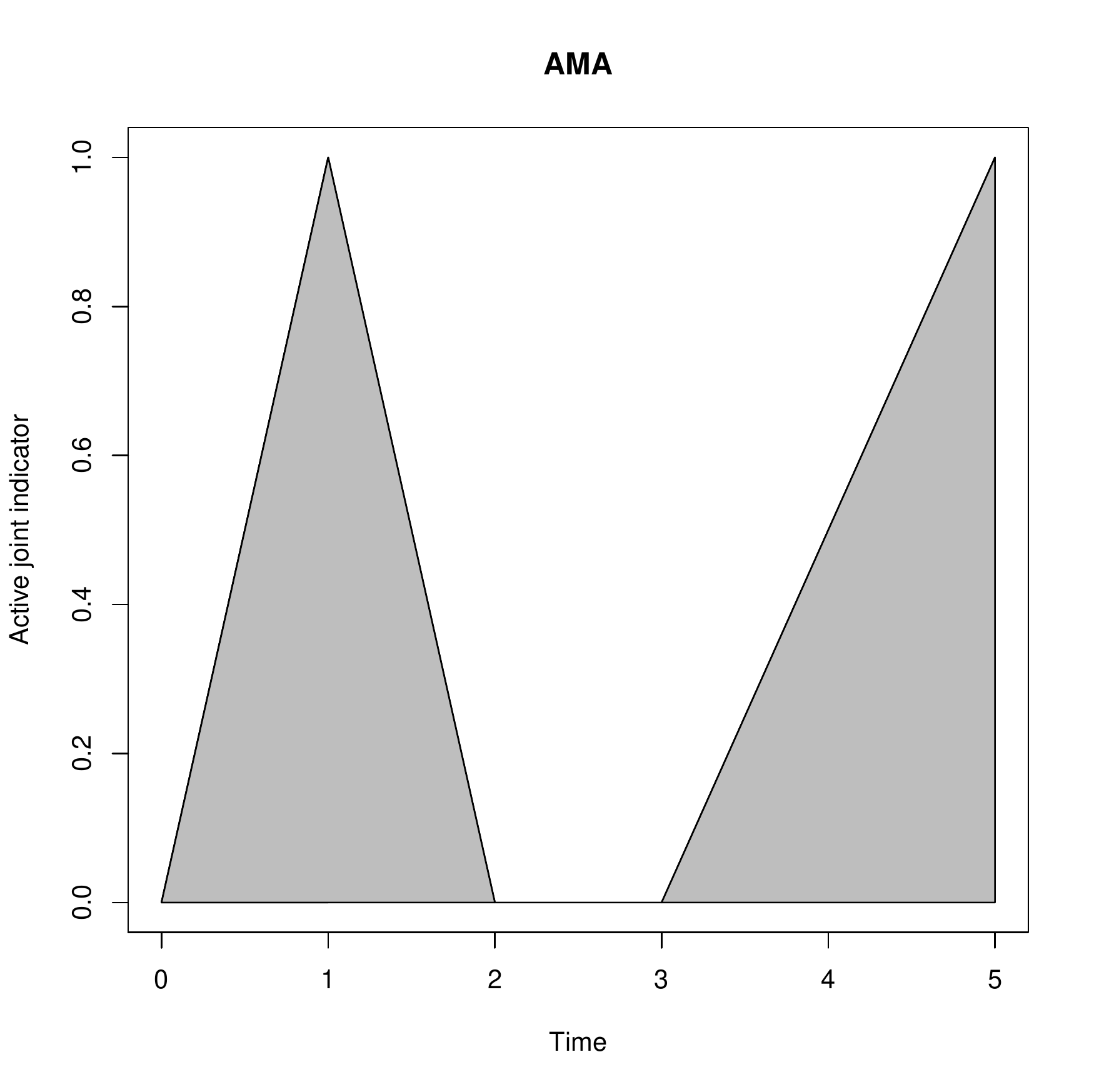}
\caption{Let $x(t)$ be realized values of $X(t)$, a binary stochastic process describing the activity process of a joint. Specifically at time $t$, $x(t)=1$ corresponds to the joint being active and $x(t)=0$ corresponds to the joint being not active. AMA$(t)$ is then calculated as $\frac{1}{t}\int_{0}^{t}x(s)\,ds$, thus resulting in a bounded measure between $[0,1]$. As $x(t)$ is intermittently observed and therefore the true path of $x(t)$ is not known, $\int_{0}^{t}x(s)\,ds$ is approximated as the area under the linearly interpolated observations of $x(t)$. For example, if a joint was observed at $t=(0,1,2,3,5)$ such that $x(t)=(0,1,0,0,1)$ respectively, then AMA$(5)$ is approximated as the gray area shaded in the figure divided by five, hence 0.4.}
\end{figure}
\indent The motivation behind the use of observation-level random effects arose from the potential need to allow for non-predictable changes in unobserved heterogeneity (Unkel \etal, 2014) with regard to the damage and activity processes. The current methodology in the literature primarily uses patient-level random effects. This forces unobserved heterogeneity to be time-invariant and we assess this assumption in our context by fitting this model (i.e. $\{U_{ij},V_{ij}\}=\{U_i,V_i\}$ $\forall j$) in addition to the proposed model. Likelihood values can then be used to informally  compare the usefulness of the proposed modelling framework (models are non-nested but contain the same number of parameters, hence the same penalty terms are obtained if information criteria such as AIC are used). In general, the estimability of random effects models, particularly variance components, will be driven by the random effects structure and the variability of the data. The observation-level random effects structure incorporates fewer shared random effects between transition intensities than its patient-level counterpart. As a consequence, there will be less variability between transition intensities containing the same random effect, thus making it harder to estimate the random effect variance when using the observation-level random effects structure. This would especially be the case when a single multi-state process is of interest, where substantial heterogeneity in the data (which could be generated by constraining transition intensities) will likely be needed for observation-level random effects models to be estimable, whilst patient-level random effects models are likely to require considerably less heterogeneity (Satten (1999) and Cook (1999) consider a single multi-state process with patient-level random effects). In a clustered multi-state process framework, heterogeneity is also generated through the differences across several multi-state processes, thus the observation-level random effects structure is far more likely to be estimable, especially as the number of processes increases. 
\subsection{Maximum likelihood estimation}
The proposed model is fitted by constructing and then maximizing the marginal likelihood. Let $X^l_i(t_{ij})$ denote the six-state process for the $l$th joint from the $i$th patient at time $t_{ij}$,  where $\{t_{i1},\ldots,t_{im_i}\}$ denotes the times of the $j$th clinic visit. Let $C_i$ be a partially observable binary variable such that $C_i=1$ with probability $1-\pi_i$ if patient $i$ is a mover (transitions are governed by the four-state model in Figure 1), and $C_i=0$ with probability $\pi_i$ otherwise (transitions are governed by the two-state model in Figure 2). The conditional likelihood contribution (condition on the dynamic covariates, $\{U_{ij}=u_{ij},V_{ij}=v_{ij}\}$ $\forall j$ and $C_i=c_i$) from the $l$th joint of the $i$th patient is then
\[\prod_{j=2}^{m_i-1}\mathbb{P}(X^l_i(t_{ij+1})=s^l_{ij+1}|X^l_i(t_{ij})=s^l_{ij};u_{ij},v_{ij},c_i)\]
where $s^l_{ij}$ represents the state corresponding to the specific combination of $(\{\bar{A},A\},\{\bar{D},D\})$ observed at $t_{ij}$ for the $l$th joint of the $i$th patient. Note that for simplicity, the likelihood contribution from the process between $t_{i1}$ and $t_{i2}$ is excluded because AMA cannot be calculated at $t_{i1}$; it requires previous observations. More details on the likelihood construction for time-homogeneous Markov models, particularly the form of the transition probabilities, can be found in Kalbfleisch and Lawless (1985). Appendix A provides the closed form transition probabilities of the six-state process. If the assumption of independence between joints from the same patient is reasonable, conditional on the random effects $U_{ij}$ and $V_{ij}$, then
\[L_i(\bm{\Theta}|C_i)=\int_{-\infty}^{\infty}\int_{-\infty}^{\infty}\prod_{l=1}^{28}\prod_{j=2}^{m_i-1}\mathbb{P}(X^l_i(t_{ij+1})=s^l_{ij+1}|X^l_i(t_{ij})=s^l_{ij};u_{ij},v_{ij},c_i)\phi(u_{ij},v_{ij};\bm{0},\Sigma)\,du_{ij}\,dv_{ij}\]
represents the likelihood contribution from the $i$th patient, still conditional on the dynamic covariates and $C_i=c_i$. Here $\bm{\Theta}$ is a vector containing all of the unknown parameters to be estimated apart from the mover-stayer probabilities $\pi_i$ and $\phi(u_{ij},v_{ij};\bm{0},\Sigma)$ denotes the zero-mean bivariate normal density with covariance matrix $\Sigma$. The overall marginal likelihood contribution from the $i$th patient is then
\[
L_i(\bm{\Theta}^*)=\left\{(1-\pi_i)L_i(\bm{\Theta}|C_i=1)\right\}^{c^{*}_i}\left\{(1-\pi_i)L_i(\bm{\Theta}|C_i=1)+\pi_i L_i(\bm{\Theta}|C_i=0)\right\}^{1-c^{*}_i}
\]
with the likelihood obtained from the product of all likelihood contributions from each patient. Here $\bm{\Theta}^*=\{\bm{\Theta},\pi_i\}$ and $c^{*}_i$ is a binary indicator such that $c^{*}_i=1$ if damaged joints are observed from patient $i$ at their last clinic visit and $c^{*}_i=0$ otherwise. The bivariate numerical integrations were computed by firstly factorizing the bivariate density function into conditional densities, i.e. $\phi(v_{ij};\rho u_{ij}\sigma_v/\sigma_u,\sigma^2_v(1-\rho^2))\phi(u_{ij};0,\sigma^2_u)$, where $\sigma^2_v$, $\sigma^2_u$ denotes the respective variance components and $\rho$ the correlation parameter, then using Gauss Hermite quadrature to evaluate each integral with respect to $u_{ij}$ and $v_{ij}$ separately. The number of quadrature points for each integration were chosen to be 15 and 30 for the observation- and patient-level random effects models respectively. Weights and nodes from the quadrature rule were then calculated using the \verb R  (2008) package \verb statmod  (2004). A sensitivity analysis indicated further quadrature points provided negligible impact on parameter estimates and log-likelihood values. The log-likelihood was maximized using the BFGS (1970) optimization routine and asymptotic standard errors for parameter estimates were obtained by evaluating and then inverting the numerically derived Hessian matrix at the maximum likelihood estimates.\\
The next subsection provides results of fitting the proposed model to the  data described in Section 2.

\subsection{Results}
In addition to the aforementioned covariates, adjustment covariates for type of joint, presence of opposite joint damage (opposite joint damaged=1 and 0 otherwise), sex (male=1 and female=0), age at arthritis onset (in years) and arthritis duration (in years) are provided through $\bm{z}^l_{ij}$. Joint type is represented through a five-level categorical variable with levels metacarpophalangeal, proximal interphalangeal, distal interphalangeal, thumb metacarpophalangeal and baseline thumb proximal interphalangeal. This covariate was included in the transition intensities associated with $\bar{A}\rightarrow A$ and $\bar{D}\rightarrow D$ with preliminary analysis demonstrating little evidence of differential recovery rates from activity (i.e. $A\rightarrow \bar{A}$). The binary variable specifying presence of opposite joint damage is motivated by previous analyses (Cresswell and Farewell, 2011 and O'Keeffe \etal , 2011) which indicate evidence of symmetric joint damage; the propensity of damage for a joint in a specific location to become damaged is increased if the contralateral joint in the other hand is earlier damaged.\\
\indent Table 1 presents the results from fitting the proposed model (with dynamic covariates and an observation-level random effects structure) to the 743 psoriatic arthritis patients described in Section 2. For comparative purposes, a model with patient-level random effects, i.e. $U_{ij}=U_i$ and $V_{ij}=V_i$, was also fitted. The results of this model and a comparison to the results of the proposed model are provided in Appendix B. The larger log-likelihood value of $-47389.11$ for the proposed model compared to the log-likelihood value of $-52784.84$ for the comparative model would suggest that preference should be given to the proposed model.
\subsubsection{Damage process}
From Table 1, it is clear that both opposite joint damage and the number of damaged joints are strongly and positively associated with an increased damage progression rate. Thus this analysis supports the results in O'Keeffe \etal (2011) concerning symmetry even after adjusting for a greater number of process features, although not adjusting for the stochastic nature of the opposite joint damage process. Activity, both current (joint is active whilst adjusting for its stochastic nature) and history (as described by AMA), are seen to be strongly and positively associated with damage progression. It is important to note that the confidence interval for the regression coefficient associated with current activity is narrower than the corresponding interval reported in O'Keeffe \etal (2011). This has likely resulted from using updated data and a more efficient estimation procedure (through dynamic covariates and random effects as opposed to a working independence assumption with a robust covariance matrix adjustment). These results therefore provide greater confidence in the strong positive association between activity and damage, and implicitly strengthens the argument that was made regarding causality.
\subsubsection{Activity process}
There is evidence that the transition intensities associated with entering and leaving the active joint state reduces once a joint has become damaged. However, as the respective (95$\%$ Wald intervals) confidence intervals contain or are close to zero, this observation must currently be regarded as suggestive. The presence of opposite joint damage and the number of damaged joints seem to be moderately/weakly associated with the activity transition intensities. In particular little association is seen with transitioning from the active joint state to the not active joint state. The strong association between history of activity and current activity transition intensities is reassuring, since the interpretation of greater amounts of previous activity increasing the transition intensity to the active state whilst decreasing the transition intensity to the not active state  is intuitive.
\subsubsection{Movers and stayers}
The percentage of stayers ($\pi\times 100\%$ where $\pi=\pi_i$ $\forall i$) was estimated to be 14$\%$ (11$\%$, 18$\%$). Empirically, when compared to the $69\%$ of patients who did not develop any damage, this estimate may seem to be a considerable underestimate of the true stayer proportion. However, because of the relationship between activity and damage, it is conceivable that many of these patients (those who did not develop damage joints) did not develop damage because they were in the not active state for long periods of continuous time, as opposed to being stayers {\it per se}. This observation is perhaps supported by Table 1 which suggests that movers have a vastly smaller transition intensity to the active state compared to stayers. A more specific investigation regarding the sojourn times of movers and stayers in the not active joint state follows in the next section. It is also worth noting that the transition intensity to the not active state is smaller for movers, however it is far less pronounced.

\begin{table}[h!]
\caption{Parameter estimates and 95$\%$ Wald intervals resulting from fitting the six-state model (described in Section 3) to 743 psoriatic arthritis patients.}
\begin{center}
\begin{tabular}{l*{6}{c}c}
\toprule
&$\bar{A}\rightarrow A$&$A\rightarrow\bar{A}$&$\bar{D}\rightarrow D$\\
\midrule
Damaged joint &-0.13 (-0.27, 0.0079)&-0.2 (-0.31, -0.088)&\\
Opposite joint & & &\\
damaged  &0.17 (0.027, 0.3)&0.09 (-0.035, 0.21)&0.83 (0.6, 1.07)\\
Attained number\\
of damaged joints &0.033 (0.021, 0.045)&-0.0039 (-0.014, 0.0061) &0.22 (0.18, 0.25)\\
Active joint & & &1.62 (1.3, 1.94)\\
AMA &2.72 (2.58, 2.87)&-0.49 (-0.61, -0.37)&2.01 (1.68, 2.34)\\
Metacarpophalangeal &0.3 (0.22, 0.372)& &-0.84 (-1.10, -0.58)\\
Proximal & & &\\
Interphalangeal &0.46 (0.38, 0.53) & &-0.15 (-0.38, 0.089)\\
Distal & & &\\
Interphalangeal &-0.18 (-0.26, -0.095) & &0.49 (0.26, 0.73)\\
Thumb\\
metacarpophalangeal &0.45 (0.36, 0.55) &&0.45 (0.17, 0.72)\\
Sex &-0.69 (-0.79, -0.59)&0.017 (-0.055, 0.088)&0.2 (-0.047, 0.44) \\
Age at arthritis\\ 
onset &0.0012 (-0.0031, 0.0055) &0.008 (0.0049, 0.011) &0.013 (0.0038, 0.023)\\
Arthritis duration &-0.021 (-0.027, -0.015) &0.0066 (0.0026, 0.011)&-0.01 (-0.023, 0.0028)\\
Stayer&1.99 (1.86, 2.12) &0.22 (0.11, 0.33) &\\
$\log(\lambda_0)$ &-3.18 (-3.4, -2.95) &0.79 (0.63, 0.94)&-9.48 (-10.08, -8.89)\\
\midrule
$\sigma^2_u$ &2.07 (1.93, 2.21) &&\\
$\alpha$ &&-0.38 (-0.42, -0.35)&\\
$\sigma^2_v$ & & &6.62 (5.89, 7.45)\\
$\rho$ &0.16 (0.1, 0.21) & &\\
$\pi$  &0.14 (0.11, 0.18)& &\\
Log-likelihood&-47389.11&&\\
\bottomrule
\end{tabular}
\end{center}
\end{table}

\section{Five-state model for mean sojourn times}
In many settings, clinical interest lies in understanding the mean sojourn times (mean amount of time a process spends in a state before a transition occurs) as a function of covariates. When two or more transitions are possible from a state of interest, the current methodology of investigating a covariate effect involves fixing other covariates at specified values (usually at their means or as a description of a particular patient) and then calculating the difference in mean sojourn times for that state by varying the covariate of interest. This methodology is implicit because under the current multi-state modelling parametrization in terms of transition intensities, direct interpretation of covariate effects on the mean sojourn times is not straightforward when two or more transitions are possible from the state of interest; the mean sojourn time is a non-linear function of covariates from different transition intensities. This section considers the novel approach of modelling the mean sojourn times directly through a model reparameterization in order to obtain easily interpretable covariate effects on this quantity. We note that this approach is possible (in our situation) because there is a smooth bijection from the transition intensities to the mean sojourn times and jump probabilities (see the next paragraph). This implies that more elaborate but computationally intensive techniques such as the use of pseudo-observations (Anderson and Perme, 2010) can be avoided.\\
\indent The specific context of interest concerns the sojourn times in the active and not active states prior to damage. Thus we revert to a three-state model for the movers by combining the activity process after damage has occurred into a single absorbing state, as depicted in Figure 5.
\begin{figure}[h!]
\centering
\begin{tikzpicture}
\tikzstyle{main}=[circle,fill=white,draw=black,text=black,minimum size=1.3cm]
\tikzstyle{connect}=[->,shorten >=1pt,auto,node distance=6cm,semithick]
  \node[main] (0)  {1 $\bar{A}$};
   \node[main] (1) [right =2cm] {2 $A$};
  \node[main] (2) [below right=1.5cm and 0.5cm of 0] {3 $D$};
\

\path (0) edge [connect, bend left] node {$\lambda_{12}$} (1);
  \path (1) edge [connect, bend left] node {$\lambda_{21}$} (0);
 \path (0) edge [pos=0.3,connect, bend right, below] node[text width=1.2cm] {$\lambda_{13}$} (2);
\path (1) edge [pos=0.3,connect, bend left] node {$\lambda_{23}$} (2);
\end{tikzpicture}
\caption{Multi-state model describing activity and damage processes jointly for movers.}
\end{figure}
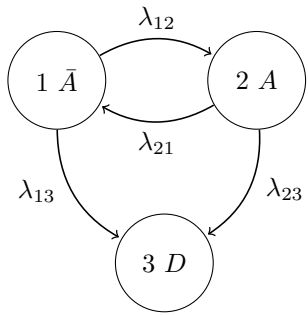
Similarly, the multi-state diagram describing possible transitions for stayers is displayed in Figure 6.\\
\begin{figure}[h!]
\centering
\begin{tikzpicture}
\tikzstyle{main}=[circle,fill=white,draw=black,text=black,minimum size=1.3cm]
\tikzstyle{connect}=[->,shorten >=1pt,node distance=10cm,semithick]
  \node[main] (0)  {4 $\bar{A}$};
  \node[main] (1) [right=2cm of 0] {5 $A$};
\
\path (0) edge [connect, bend left] node [pos=0.5, sloped, above]{$\lambda_{45}$} (1);
  \path (1) edge [connect, bend left] node [pos=0.5, sloped, below]{$\lambda_{54}$} (0);
\end{tikzpicture}
\caption{
Multi-state model describing activity process for stayers.}
\end{figure}
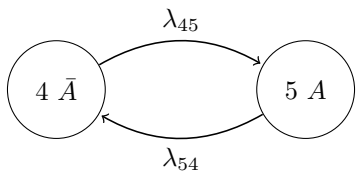

\indent Whilst continuous-time Markov processes can be viewed in terms of transition intensities specifying risks of transitioning through states (as thought of previously), another natural way to view such processes is in terms of sojourn times and jump probabilities being associated with each state. As mentioned, the sojourn time in a state describes the amount of time the process spends in that state before a transition occurs. Then at the point of transitioning, the jump probabilities informs the multinomial distribution of the possible set of states in which the process will jump to next. For the process we consider (a continuous-time Markov process), the sojourn time in state $r$ can be shown to have an exponential distribution with mean $1/\sum_{k\neq r}\lambda_{rk}$ and jump probabilities, i.e. $\mathbb{P}(\text{jumps to state}\ k|\ \text{jumps from state}\ r)=\lambda_{rk}/\sum_{k\neq r}\lambda_{rk}$. See pages $259-260$ in Grimmett and Stirzaker (2001) for more details. To again more simply investigate the difference in the activity process between movers and stayers, we parametrize as follows:
\begin{align*}
\mu^l_{4ij}&=\mu^l_{1ij}\exp(\beta^{\bar{A}}_{Stayer})\\
\mu^l_{5ij}&=\mu^l_{2ij}\exp(\beta^{\bar{A}}_{Stayer})
\end{align*}
where $\mu^l_{rij}$ represents the mean sojourn time in state $r$ for the $l$th joint of the $i$th patient at time $t_{ij}$. Regression models for the mean sojourn times and jump probabilities can then be specified as follows:
\begin{align*}\tag{3}
\mu^{l}_{1ij}&=\mu^{\bar{A}}_0\exp(\bm{\beta}^{\bar{A}} \bm{z}^l_{ij}+u_{ij})\\
\mu^{l}_{2ij}&=\mu^{A}_0\exp(\bm{\beta}^{A} \bm{z}^l_{ij}+\alpha_1 u_{ij})\\
p^{l}_{13ij}/(1-p^{l}_{13ij})&=p^{\bar{A}D}_0\exp(\bm{\beta}^{\bar{A}D} \bm{z}^l_{ij}+v_{ij})\\
p^{l}_{23ij}/(1-p^{l}_{23ij})&=p^{AD}_0\exp(\bm{\beta}^{AD} \bm{z}^l_{ij}+\alpha_2 v_{ij})\\
\end{align*}
where $p^{l}_{\bar{A}Dij}$ and $p^{l}_{ADij}$ denotes the jump probabilities from state $\bar{A}\rightarrow D$ and $A\rightarrow D$ respectively for the $l$th joint of the $i$th patient at $t_{ij}$. The right hand side of each regression equation in (3) contains a baseline (indicated by 0 in the subscript) multiplied by the exponent of the sum of a linear predictor and linear function of realizations of random effects, as before. Dynamic covariates (AMA and attained number of damaged joints) and random effects are again included to reflect features of the processes as described in Section 3. The random effects follows a zero mean bivariate normal distribution and are independent across time.\\
\indent The model fitting procedure follows from Section 3.1 after having specified the transition probabilities which as a function of transition intensities can be found in Appendix C. Thus the following set of equations completes the procedure:
\begin{align*}
\lambda_{12}&=(1-p_{13})/\mu_{1}\\
\lambda_{13}&=p_{13}/\mu_{1}\\
\lambda_{21}&=(1-p_{23})/\mu_{2}\\
\lambda_{23}&=p_{23}/\mu_{2}\\
\lambda_{45}&=1/\mu_{4}\\
\lambda_{54}&=1/\mu_{5}.
\end{align*}
\section{Results}
Along with dynamic covariates, the presence of opposite joint damage, sex, age at arthritis onset and arthritis duration was included in the analysis, as before. Table 2 presents the results from fitting the five-state model described in Section 4 to the 743 psoriatic arthritis patients described in Section 2. From the table, the presence of opposite joint damage provides a slight increase in the mean sojourn time in the not active joint state and greatly increases the probability of directly transitioning to damage (as opposed to active and not damaged) once a transition occurs. However, when a joint is active, there is little evidence to suggest opposite joint damage influences the sojourn times nor the next state probability. These results would seem to indicate that the presence of opposite joint damage is particularly relevant when a joint is not active. A  large number of damaged joints, whilst substantially increasing the jump probabilities to damage as opposed to active/not active respectively, provides little impact on the mean sojourn times in the active and not active states. As expected, greater amounts of previous activity (as described by AMA) decreases the mean sojourn time in the not active state and increases the mean sojourn time in the active state. The table also suggests AMA is strongly and positively associated with the jump probability to damage when in the active state but not in the not active state. Thus current activity is strengthened by the history of activity when dictating the next state of the process, but jumping to damage or active from the not active state could be unrelated to the history of activity. As hypothesized in the previous section, stayers have a far shorter mean sojourn time in the not active state, whilst a slightly shorter sojourn time in the active state.\\
\indent Table 4 in Appendix D provides the results of fitting the five state model with a patient-level random effects structure (i.e. $U_{ij}=U_i$ and $V_{ij}=V_i$ $\forall j$) to the data described in Section 2. The resulting log-likelihood value was -49256.31, which is far smaller then -44346.26 obtained from the proposed model. In this context, the observation-level random effects structure, after including dynamic covariates, again seems to be the more appropriate random effects structure for the data. 

\begin{sidewaystable}[ph!]
\caption{Parameter estimates and 95$\%$ Wald intervals resulting from fitting the five-state model (described in Section 4) to 743 psoriatic arthritis patients.}
\begin{center}
\begin{tabular}{l*{6}{c}c}
\toprule
&Sojourn times & &Jump probabilities &\\
\midrule
&$\bar{A}$&$A$&$\bar{A}\rightarrow D$&$A\rightarrow D$\\
\midrule
Opposite joint && & &\\
damaged &-0.28 (-0.44, -0.12) &-0.1 (-0.27, 0.061) &1.6 (0.87, 2.34) &0.54 (-0.18, 1.26)\\
Attained number\\
of damaged joints &-0.09 (-0.11, -0.074) &0.0075 (-0.0057, 0.021) &0.21 (0.14, 0.29) &0.16 (0.11, 0.22)\\
AMA &-2.98 (-3.13, -2.83)&0.39 (0.26, 0.52) &-0.25 (-1.02, 0.52) &0.84 (0.21, 1.48)\\
Sex &0.68 (0.58, 0.78)&0.0053 (-0.067, 0.078) &1.24 (0.74, 1.74) &0.43 (-0.022, 0.89)\\
Age at arthritis\\ 
onset &-0.0021 (-0.0064, 0.0023) &-0.0078 (-0.011, -0.0047) &1.24 (0.74, 1.74) &0.43 (-0.022, 0.89)\\
Arthritis duration &0.019 (0.014, 0.025)&-0.0036 (-0.0076, 0.00047) &0.037 (0.016, 0.059) &-0.023 (-0.051, 0.0053) \\
Stayer &-1.93 (-2.06, -1.8) &-0.17 (-0.29, -0.049) & &\\
$\log(\mu_0)$&3.04 (2.82, 3.25) &-0.83 (-0.99, -0.67) & &\\
$\log(p_0)$ & & &-9.27 (-10.9, -7.65) & -5.68 (-6.81, -4.55)\\
\midrule
$\sigma^2_u$ &2.05 (1.92, 2.2)  & & &\\
$\alpha_1$ &&-0.35 (-0.39, -0.32)& &\\
$\sigma^2_v$ &  & &17.4 (12.66, 23.92) &\\
$\alpha_2$ &&& &0.44 (0.32, 0.56)\\
$\rho$ &0.0041 (-0.048, 0.056) & & &\\
$\pi$ &0.15 (0.12, 0.18) & & &\\
Log-likelihood &-44346.26 & & &\\
\bottomrule
\end{tabular}
\end{center}
\end{sidewaystable}

\section{Discussion}
This research was motivated by reproducing prior results and undertaking new investigations into disease course and progression. For this purpose, a single unifying clustered multi-state modelling framework which allows simultaneous investigations of multiple clinical aspects was proposed. The results obtained therefore yield greater confidence when compared with multiple univariate investigations, which were performed previously, since they are based on adjusting for other important process characteristics. From a clinical perspective, the relationship between activity and damage was demonstrated as pronounced since both history and current activity were positively related to damage progression and jumping to damage once a joint  immediately leaves the active state. In terms of the reverse relationship, damage onset is seen to slow the activity process although the confidence intervals for the relevant regression coefficients indicates no change is a distinct possibility, maybe due to the far fewer observed transitions after damaged has occurred. Interestingly, both models seem to identify a subpopulation of approximately $15\%$ who are rapidly fluctuating in their activity process, yet are at minimal risk of damage, perhaps because they have shorter sojourn times in the active joint state. An avenue of future work could involve identifying these patients especially because their treatment strategies should conceivably not consist of potent drugs, which may cause unpleasant side effects, but soft drugs to reduce joint swelling and pain. It is also reassuring that both models do not contradict any well held clinical beliefs.\\
\indent From a statistical point of view, the novel aspects of this research includes the proposing of an observation-level random effects structure combined with dynamic covariates, a mover-stayer structure whereby movers and stayers can have different effects on transition intensities in which they are not implicitly defined for, and a model parameterization which allows easily interpretable covariate effects to act on the sojourn times and jump probabilities. In our context, the usefulness of the proposed methodology was demonstrated through new clinical insights and substantial improvements in likelihood values over the use of standard methodology (patient-level random effect models). Whilst the proposed methodology was described in terms of specific, but fairly complex, six- and five-state models for the margins, extensions to general clustered continuous time Markov models are straightforward. In particular, the proposed model parameterization in terms of sojourn times and jump probabilities, and mover-stayer effects on transition intensities are also applicable to univariate Markov multi-state processes and therefore can provide a useful framework for inference in many clinical settings.\\
\indent Overall, this research represents our efforts to provide a comprehensive investigation into many clinical aspects of interest at the finest level of detail (individual joint-level). Although there are foreseeable model extensions, it is important to bear in mind the computationally intensive nature of fitting clustered multi-state models with random efforts, especially when reversible multi-state models are involved (transitions to and fro states exist). Some examples of potentially more appropriate extensions could include relaxing the time-homogeneous assumption beyond adjusting for arthritis duration, relating previous history to current transitions through more accurate measures then the proposed dynamic covariates and dividing the damage onset state into various states of damage severity. Such extensions, as with many others, will usually require transitioning from piecewise constant approximations to reflecting the true stochastic nature of the outcome, covariates and latent processes, which has been seen as one of the main drivers of complexifying the model fitting procedure, due to the larger number of integrations/differential equations required to be computed/solved. Nevertheless, as demonstrated here, it is important to identify and provide adjustments for important process characteristics where possible, in which model comparison is useful. With respect to understanding clinical aspects of psoriatic arthritis, this has provided new knowledge and greater confidence in prior results on less general methodology.

\begin{appendices}
\section{}
The transition probabilities, i.e. $p_{rs}(t)\equiv p_{rs}(t;\bm{\lambda})=\mathbb{P}(X(t+s)=s|X(s)=r;\bm{\lambda})$ where $\bm{\lambda}$ is a vector of the required transition intensities, for the four- and two-state process depicted in Figure 2 and 3 respectively can be calculated through taking the matrix exponential of the relevant transition intensity matrix. Specifically, let 
\[Q_1=\bordermatrix{&\cr
                  &-\lambda_{12}-\lambda_{13} &\lambda_{12}&\lambda_{13}&0  \cr
           &\lambda_{21}&-\lambda_{21}-\lambda_{24}&0&\lambda_{24} \cr
&0&0&-\lambda_{34}&\lambda_{34}\cr
&0&0&\lambda_{43}&-\lambda_{43}},\]
then the transition probabilities for the four-state process can be obtained from the $(r,s)$th entry of $\exp(Q_1t)$. We use Mathematica to compute the following matrix exponential and this results in
\begin{align*}
p_{11}(t)&=\frac{\exp(-\Lambda t/2)}{2\gamma}[(\Lambda_1+\gamma)\exp(-\gamma t/2)+(\gamma-\Lambda_1)\exp(\gamma t/2)]\\
p_{12}(t)&=\frac{\lambda_{12}}{\gamma_1}[\exp(-(\Lambda-\gamma_1)t/2)-\exp(-(\Lambda+\gamma_1)t/2)]\\
p_{13}(t)&=\frac{\lambda_{43}}{\Lambda_2}+\frac{\exp(-\Lambda_2t)}{\Lambda_2}\frac{[\lambda_{13}\lambda_{34}(\lambda_{21}+\lambda_{24}-\Lambda_2)-\lambda_{12}\lambda_{24}\lambda_{43}]}{[(\lambda_{21}+\lambda_{24}-\Lambda_2)(\lambda_{13}-\Lambda_2)-\lambda_{12}(\lambda_{24}-\Lambda_2)]}\\
&+\lambda_{13}\exp(-(\Lambda-\gamma)t/2)\frac{[\lambda_{12}\lambda_{21}+(\lambda_{12}+\lambda_{13})(\Lambda_1-\gamma)/2+\lambda_{43}(-\Lambda_1+\gamma-2\lambda_{34})/2+\gamma_2]}{[-(\Lambda-\gamma)^3/2+3(\Lambda-\gamma)^2(\Lambda+\Lambda_2)/4-(\Lambda-\gamma)\gamma_3+\gamma_4]}\\
&+\lambda_{13}\exp(-(\Lambda+\gamma)t/2)\frac{[\lambda_{12}\lambda_{21}+(\lambda_{12}+\lambda_{13})(\Lambda_1+\gamma)/2+\lambda_{43}(-\Lambda_1-\gamma-2\lambda_{34})/2+\gamma_2]}{[-(\Lambda+\gamma)^3/2+3(\Lambda+\gamma)^2(\Lambda+\Lambda_2)/4-(\Lambda+\gamma)\gamma_3+\gamma_4]}\\
p_{14}(t)&=1-p_{11}(t)-p_{12}(t)-p_{13}(t)
\end{align*}
where
\begin{align*}
\Lambda&=\lambda_{12}+\lambda_{13}+\lambda_{21}+\lambda_{24}\\
\Lambda_1&=\lambda_{12}+\lambda_{13}-\lambda_{21}-\lambda_{24}\\
\Lambda_2&=\lambda_{34}+\lambda_{43}\\
\gamma&=\sqrt{\Lambda^2-4(\lambda_{12}\lambda_{24}+\lambda_{13}(\lambda_{21}+\lambda_{24}))}\\
\gamma_1&=\sqrt{\lambda^2_{12}+2\lambda_{12}(\lambda_{13}+\lambda_{21}-\lambda_{24})+(-\lambda_{13}+\lambda_{21}+\lambda_{24})^2}\\
\gamma_2&=\frac{\lambda_{43}}{\lambda_{13}}[\lambda_{12}\lambda_{24}+\lambda_{13}\lambda_{34}]\\
\gamma_3&=\Lambda_2(\lambda_{21}+\lambda_{24})+\lambda_{13}\lambda_{21}+(\lambda_{12}+\lambda_{13})(\lambda_{24}+\Lambda_2)\\
\gamma_4&=\Lambda_2(\lambda_{13}\lambda_{21}+\lambda_{24}(\lambda_{12}+\lambda_{13})).
\end{align*}
As the four-state process is symmetric, it is easy to verify that
\begin{align*}
p_{21}(t;\lambda_{12},\lambda_{13},\lambda_{21},\lambda_{24},\lambda_{34},\lambda_{43})&=p_{12}(t;\lambda_{21},\lambda_{24},\lambda_{12},\lambda_{13},\lambda_{43},\lambda_{34})\\
p_{22}(t;\lambda_{12},\lambda_{13},\lambda_{21},\lambda_{24},\lambda_{34},\lambda_{43})&=p_{11}(t;\lambda_{21},\lambda_{24},\lambda_{12},\lambda_{13},\lambda_{43},\lambda_{34})\\
p_{23}(t;\lambda_{12},\lambda_{13},\lambda_{21},\lambda_{24},\lambda_{34},\lambda_{43})&=p_{14}(t;\lambda_{21},\lambda_{24},\lambda_{12},\lambda_{13},\lambda_{43},\lambda_{34})\\
p_{24}(t;\lambda_{12},\lambda_{13},\lambda_{21},\lambda_{24},\lambda_{34},\lambda_{43})&=p_{13}(t;\lambda_{21},\lambda_{24},\lambda_{12},\lambda_{13},\lambda_{43},\lambda_{34}).\\
\end{align*}
Finally, we have
\begin{align*}
p_{33}(t)&=1-p_{34}(t)\\
p_{34}(t)&=\lambda_{34}(1-\exp(-\Lambda_2t))/\Lambda_2\\
p_{43}(t)&=\lambda_{43}(1-\exp(-\Lambda_2t))/\Lambda_2\\
p_{44}(t)&=1-p_{34}(t).
\end{align*}
Similarly for the two-state process we have
\begin{align*}
p_{55}(t)&=1-p_{56}(t)\\
p_{56}(t)&=\lambda_{56}(1-\exp(-\Lambda_3t))/\Lambda_3\\
p_{65}(t)&=\lambda_{65}(1-\exp(-\Lambda_3t))/\Lambda_3\\
p_{66}(t)&=1-p_{56}(t).
\end{align*}
where $\Lambda_3=\lambda_{56}+\lambda_{65}$.

\section{}
Table 3 provides the results from fitting the six-state model with a patient-level random effects structure to the 743 psoriatic arthritis patients described in Section 2. Rather reassuringly, most regression coefficients of primary interest, including the stayer proportion estimate, from Table 3 are seen to have similar interpretations to those obtained from Table 1. The regression coefficients associated with the dynamic covariates AMA and attained number of damaged joints have however resulted in markedly different estimates than those seen in Table 1. Specifically, the effects of the dynamic covariates are greatly attenuated, and furthermore, the attained number of damaged joints is now seen to have a possibly counter-intuitive negative association with damage progression. Both dynamic covariates and patient-level random effects adjust for a patient's propensity of gaining/recovering from activity and gaining damage, thus are likely to be confounded when introduced simultaneously in the model. The regression coefficients associated with the dynamic covariates are now perhaps more difficult to interpret than before. Aalen \etal (2008) provides a discussion on the relationship between dynamic models (with dynamic covariates) and frailty models (with patient-level random effects). 
 \begin{table}[h!]
\caption{Parameter estimates and 95$\%$ Wald intervals resulting from fitting the six-state model with a patient-level random effects structure to 743 psoriatic arthritis patients.}
\begin{center}
\begin{tabular}{l*{6}{c}c}
\toprule
&$\bar{A}\rightarrow A$&$A\rightarrow\bar{A}$&$\bar{D}\rightarrow D$\\
\midrule
Damaged joint &-0.046 (-0.18, 0.086)&-0.18 (-0.27, -0.069)&\\
Opposite joint & & &\\
damaged  &0.13 (-0.0031, 0.27)&0.087 (-0.037, 0.21)&0.72 (0.52, 0.92)\\
Attained number\\
of damaged joints &-0.04 (-0.049, -0.03)&0.022 (0.013, 0.031) &-0.036 (-0.055, -0.018)\\
Active joint & & &1.32 (1.03, 1.61)\\
AMA &1.8 (1.68, 1.93)&-0.34 (-0.45, -0.23)&1.7 (1.39, 2)\\
Metacarpophalangeal &0.23 (0.16, 0.3)& &-0.81 (-1.1, -0.57)\\
Proximal & & &\\
Interphalangeal &0.36 (0.29, 0.43) & &-0.18 (-0.4, 0.049)\\
Distal & & &\\
Interphalangeal &-0.19 (-0.27, -0.12)& &0.36 (0.14, 0.58)\\
Thumb\\
metacarpophalangeal &0.39 (0.3, 0.48) &&0.36 (0.095, 0.62)\\
Sex &-0.73 (-0.8, -0.66)&0.011 (-0.051, 0.073)&-0.2 (-0.45, 0.042) \\
Age at arthritis\\ 
onset &0.014 (0.011, 0.018) &0.0042 (0.0015, 0.0069) &-0.012 (-0.024, -0.00034)\\
Arthritis duration &-0.037 (-0.04, -0.033) &0.0058 (0.0022, 0.0095)&0.057 (0.047, 0.067)\\
Stayer&2.25 (2.12, 2.37) &0.37 (0.26, 0.47) &\\
$\log(\lambda_0)$ &-2.88 (-3.07, -2.69) &0.8 (0.66, 0.94)&-7 (-7.55, -6.45)\\
\midrule
$\sigma^2_u$ &1.33 (1.21, 1.46) &&\\
$\alpha$ &&-0.23 (-0.28, -0.19)&\\
$\sigma^2_v$ & & &3.19 (2.73, 3.73)\\
$\rho$ &0.28 (0.2, 0.35) & &\\
$\pi$  &0.17 (0.13, 0.21)& &\\
Log-likelihood&-52784.84&&\\
\bottomrule
\end{tabular}
\end{center}
\end{table}

\section{}
The transition probabilities of the three-state model were again calculated using the matrix exponential function in Mathematica. This resulted in
\begin{align*}
p_{11}(t)&=\frac{x_1\exp(r_1t)-x_2\exp(r_2t)}{x_1-x_2}\\
p_{12}(t)&=\frac{x_1x_2(\exp(r_2t)-\exp(r_1t))}{x_1-x_2}\\
p_{13}(t)&=1-p_{11}(t)-p_{12}(t)\\
p_{21}(t)&=\frac{\exp(r_1t)-\exp(r_2t)}{x_1-x_2}\\
p_{22}(t)&=\frac{x_1\exp(r_2t)-x_2\exp(r_1t)}{x_1-x_2}\\
p_{23}(t)&=1-p_{21}(t)-p_{22}(t)\\
p_{3j}(t)&=1_{j=3}\\
\end{align*}
where $1_{j=3}$ is an indicator function taking the value $1$ when $j=3$ and $0$ otherwise, and
\begin{align*}
r_1&=\frac{\lambda_{11}+\lambda_{22}+\sqrt{(\lambda_{11}-\lambda_{22})^2+4\lambda_{12}\lambda_{21}}}{2}\\
r_2&=\frac{\lambda_{11}+\lambda_{22}-\sqrt{(\lambda_{11}-\lambda_{22})^2+4\lambda_{12}\lambda_{21}}}{2}\\
x_j&=\frac{r_j-\lambda_{22}}{\lambda_{21}}.\\
\end{align*}
Here $\lambda_{11}=-\lambda_{12}-\lambda_{13}$ and $\lambda_{22}=-\lambda_{21}-\lambda_{23}$.

\section{}
\begin{sidewaystable}[h!]
\caption{Parameter estimates and 95$\%$ Wald intervals resulting from fitting the five-state model with a patient-level random effects structure to 743 psoriatic arthritis patients.}
\begin{center}
\begin{tabular}{l*{6}{c}c}
\toprule
&Sojourn times& &Jump probabilities&&\\
\midrule
&$\bar{A}$&$A$&$\bar{A}\rightarrow D$&$A\rightarrow D$\\
\midrule
Opposite joint && & &\\
damaged &-0.28 (-0.44, -0.11) &-0.19 (-0.35, -0.026) &0.95 (0.44, 1.46) &0.44 (0.01, 0.86)\\
Attained number\\
of damaged joints &0.012 (-0.0022, 0.026) &-0.033 (-0.044, -0.022) &-0.3 (-0.45, -0.14)&0.056 (0.019, 0.093)\\
AMA &-1.98 (-2.11, -1.85)&0.29 (0.17, 0.4) &0.7 (0.068, 1.32) &0.87 (0.34, 1.4)\\
Sex &0.79 (0.7, 0.88)&0.012 (-0.054, 0.078) &1.14 (0.51, 1.77) &0.36 (-0.066, 0.79)\\
Age at arthritis\\ 
onset &-0.014 (-0.019, -0.0098) &-0.0028 (-0.0054, -0.00025) &0.032 (0.0095, 0.054) &-0.0082 (-0.03, 0.013)\\
Arthritis duration &0.036 (0.032, 0.039)&-0.0024 (-0.0062, 0.0013) &0.18 (0.13, 0.23) &0.041 (0.017, 0.065) \\
Stayer &-2.23 (-2.36, -2.1) &-0.35 (-0.46, -0.23) & &\\
$\log(\mu_0)$&2.71 (2.49, 2.94) &-0.91 (-1.04, -0.77) & &\\
$\log(P_0)$ & & &-8.8 (-10.43, -7.18) &-5.61 (-6.59, -4.62) \\
\midrule
$\sigma^2_u$ &1.28 (1.16, 1.41) & & &\\
$\alpha_1$ &&-0.24 (-0.29, -0.19)& &\\
$\sigma^2_v$ &  & &6.96 (4.41, 10.98) &\\
$\alpha_2$ &&& &0.63 (0.45, 0.81)\\
$\rho$ &-0.057 (-0.15, 0.032) & & &\\
$\pi$ &0.16 (0.13, 0.2 & & &\\
Log-likelihood &-49256.31 & & &\\
\bottomrule
\end{tabular}
\end{center}
\end{sidewaystable}

\section*{Acknowledgments}
This research was financially supported by grants from the UK Medical Research Council [Unit program numbers U105261167 and MC\_UP\_1302/3]. We also acknowledge the patients in the Toronto psoriatic arthritis clinic.

\end{appendices}
\clearpage

\def\bibindent{1em}

\end{document}